\documentclass[a4paper,11pt]{article}
\pdfoutput=1 
\usepackage[utf8]{inputenc}
\usepackage{jheppub} 
\usepackage{amsmath}
\usepackage{bbold}
\usepackage{xcolor}
\usepackage{subfig}
\usepackage{graphicx}
\usepackage{hyperref}
\newcommand{\an}[1]{\langle#1\rangle}

\newcommand{\mAn}{\mathcal{A}_n}
\newcommand{\D}{\mathrm{d}}
\newcommand{\p}{\partial}
\newcommand{\calA}{{\cal A}}

\newcommand{\calK}{{\cal K}}
\newcommand{\calO}{{\cal O}}

\newcommand{\calS}{{\cal S}}

\def\pa{\partial}
\def\a{\alpha}
\def\b{\beta}

\def\d{\delta}

\def\s{\sigma}

\title{Celestial Dual Superconformal Symmetry, 

MHV Amplitudes and Differential Equations }
\author[a,b]{Yangrui Hu,}
\author[a]{Lecheng Ren,}
\author[a]{Akshay Yelleshpur Srikant,}
\author[a]{Anastasia Volovich}

\affiliation[a]{Department of Physics,
Brown University,
Providence, RI 02912, USA}
\affiliation[b]{Brown Theoretical Physics Center,
Box S, 340 Brook Street, Barus Hall,
Providence, RI 02912, USA}

\abstract{Celestial and momentum space amplitudes for massless particles are related to each other by a change of basis provided by the Mellin transform. Therefore properties of celestial amplitudes have counterparts in momentum space amplitudes and vice versa. In this paper, we study the celestial avatar of dual superconformal symmetry of $\mathcal{N}=4$ Yang-Mills theory. We also analyze various differential equations known to be satisfied by celestial $n$-point tree-level MHV amplitudes and identify their momentum space origins. }

\begin{document}
\maketitle
\section{Introduction}
The quest for flat space holography has recently received a boost owing to the realization \cite{He:2015zea, Kapec:2016jld,  Bagchi:2016bcd,
Cheung:2016iub, Pasterski:2016qvg, Pasterski:2017ylz, Cardona:2017keg, Ball:2019atb} that scattering amplitudes in 4D flat spacetime can be recast as correlation functions of a 2D conformal field theory living on the celestial sphere. The appearance of the 2D conformal symmetry is due to the fact that the Lorentz group SL$(2, \mathbb{C})$ acting at null infinity of 4D Minkowski spacetime is mapped to the global conformal group acting on the celestial sphere. Such a theory, dubbed celestial CFT (CCFT), is a potential candidate for a holographic description of the flat space S-matrix. Unsurprisingly, it has been the subject of intense study for the past few years. However, our understanding of CCFTs remains primitive, and in particular, a dictionary similar to the one for the AdS/CFT correspondence remains elusive. \\

A path towards a better understanding of CCFTs can be forged by generating more ``data". This involves translating well understood aspects of momentum space amplitudes into statements about celestial correlators, as well as mapping momentum space amplitudes onto the celestial sphere providing us with explicit examples of celestial amplitudes. The last few years have seen a lot of progress on both accounts. Soft and collinear theorems have been studied in \cite{Donnay:2018neh, Himwich:2019dug, Fan:2019emx, Pate:2019mfs, Adamo:2019ipt, Nandan:2019jas, Puhm:2019zbl, Guevara:2019ypd, Fotopoulos:2019vac, Fotopoulos:2020bqj, Himwich:2020rro}. The factorization of gauge theory amplitudes into ``hard" and ``soft/collinear" factors has been extended to and interpreted on the celestial sphere \cite{Magnea:2021fvy, Gonzalez:2021dxw, Arkani-Hamed:2020gyp}. Attempts have been made to demonstrate the existence of celestial double copy relations \cite{Casali:2020uvr, Casali:2020vuy}. Several momentum space amplitudes have been mapped onto the celestial sphere including tree-level gluon amplitudes \cite{Pasterski:2017ylz, Schreiber_2018}, one-loop amplitudes in scalar \cite{Banerjee:2017jeg} and gauge theories \cite{Albayrak:2020saa}, all-loop four-point amplitudes \cite{Gonzalez:2020tpi} and string amplitudes \cite{Stieberger:2018edy}. A parallel path towards the same goal is to leverage symmetries to directly obtain constraints on CCFTs along the lines of \cite{Banerjee:2020kaa, Banerjee:2020vnt, Banerjee:2020zlg, Pate:2019lpp, Fan:2021isc, Strominger:2021lvk, Atanasov:2021cje, Guevara:2021abz, Crawley:2021ivb, Atanasov:2021oyu}.   \\

Theories with a large symmetry group are amenable to both lines of attack. 
Recently celestial superconformal symmetry of $\mathcal{N}=4$ Yang-Mills has been studied in \cite{Brandhuber:2021nez, Jiang:2021xzy}.  
In  this paper, we 
study the celestial avatar of the dual superconformal symmetry \cite{Drummond:2008vq}. Celestial $n$-point tree-level MHV amplitudes have been computed in \cite{Schreiber_2018} and found to be Aomoto-Gelfand hypergeometric functions, which satisfy a known set of differential equations \cite{aomoto}. We identify these equations as arising from momentum conservation and $GL(n-4)$ transformations. Another set of differential equations for the full MHV amplitude has been found in \cite{Banerjee:2020vnt}. We derive versions of these equations satisfied by colour ordered amplitudes and generalize them. We rewrite them in momentum space and identify them as encoding properties of amplitudes under BCFW shifts. Tree-level amplitudes in other helicity sectors and loop amplitudes are generalized hypergeometric functions satisfying more complex differential equations. Celestial graviton amplitudes are also known to satisfy differential equations \cite{Banerjee:2020zlg}. While we have restricted ourselves to MHV gluon amplitudes in this work, we hope to return to these topics in the future \cite{celestiahedron}.\\

This paper is organized as follows. In Section~\ref{sec:celestialsup}, we quickly review superamplitudes and their celestial counterparts. We then discuss dual conformal symmetry of momentum space amplitudes in Section~\ref{sec:reviewdc} and present its celestial counterpart in Section~\ref{sec:celestialdc2}. In Section~\ref{sec:BGeqs}, we derive momentum space generalizations of the differential equations found in \cite{Banerjee:2020vnt} by connecting them to the behaviour of amplitudes under BCFW shifts and in Section~\ref{sec:hyperpde}, we provide physical interpretations for the hypergeometric equations satisfied by the celestial MHV tree-level amplitudes. Finally, in Section~\ref{sec:relation}, we discuss the relation between these differential equations.

\section{Superamplitudes and celestial superamplitudes}
\label{sec:celestialsup}
Amplitudes in supersymmetric theories, particularly in planar $\mathcal{N} = 4$ Yang-Mills have been the focus of a lot of recent work. The field content of $\mathcal{N}=4$ Yang-Mills can be packaged into an on-shell superfield defined by 
\begin{align}
    \label{eq:onshellsf}
    \Psi (p_i) =& G^{+} (p_i) + \eta_i^A \Gamma_A (p_i) + \frac{1}{2!} \eta_i^A \eta_i^B \Phi_{AB} (p_i) + \frac{1}{3!} \epsilon_{ABCD}\eta_i^A \eta_i^B \eta_i^C \bar{\Gamma}^D  (p_i) \\
    \nonumber &+ \frac{1}{4!} \epsilon_{ABCD} \eta_i^A \eta_i^B \eta_i^C \eta_i^D G^-  (p_i)~~.
\end{align}
Here $G^\pm$ are the $\pm$ helicity gluons, $\Gamma^D$ are the gluinos and $\Phi_{AB}$ are the scalars. $A, B, \dots$ are $SU(4)$ R-symmetry indices. For more details, refer to \cite{Elvang:2015rqa} and \cite{Henn:2014yza}. The natural states to scatter in this theory are $\Psi (p_i)$ and the corresponding scattering amplitudes  are called superamplitudes $\calA_n$\footnote{We will focus on the colour ordered superamplitudes in the planar limit and denote them by $\calA_n$. An ordering will be specified only when necessary.}. These amplitudes can be expanded as polynomials in the Grassmann variables $\eta_i$ as
\begin{align}
    \label{eq:MHVdecomp}
    \calA_n = \sum_{k=0}^{n-4} \calA_{n,k}~~,
\end{align}
where $\calA_{n,k}$ is the N$^k$MHV amplitude and has $4(k+2)$ Grassmann variables. The tree-level MHV amplitude can be compactly written as   
\begin{equation}
    {\cal A}_{n,0} ~=~ \frac{\delta^{(8)} \left(Q\right)}{\langle 12\rangle\langle 23 \rangle\cdots \langle n1 \rangle } \delta^{(4)} \left( \sum_{i=1}^n p_i \right) ~=~ \frac{\frac{1}{2^4}\prod_{A=1}^4\sum_{i,j=1}^n\langle ij \rangle \eta_i^A\eta_j^A}{\langle 12\rangle\langle 23 \rangle\cdots \langle n1 \rangle } \delta^{(4)} \left( \sum_{i=1}^n p_i \right).
    \label{eq:superamplitude}
\end{equation}
Here $\lambda_i, \lambda_j$ are the usual spinor helicity variables and they appear in (\ref{eq:superamplitude}) in the following Lorentz invariant combinations,
\begin{align}
    \label{eq:squareanglebrackets}
    \an{ij} = \epsilon_{\alpha \, \beta} \lambda_i^\alpha \, \lambda_j^\beta \qquad,\qquad [ij] = -\epsilon_{\dot{\alpha} \, \dot{\beta}} \tilde{\lambda}_i^{\dot{\alpha}} \, \tilde{\lambda}_j^{\dot{\beta}}~~.
\end{align}
The MHV gluon amplitude\footnote{We will denote superamplitudes by ${\cal A}_{n,k}$ and the MHV gluon amplitude by ${\cal M}_{n}$.} (with particles $s$ and $t$ having negative helicity and the remaining $n-2$ having positive helicity), which we denote by $\mathcal{M}_n$, is contained in (\ref{eq:superamplitude}) as the coefficient of $(\eta_s)^4 (\eta_t)^4$
\begin{align}
    \label{eq:n-gluonamp}
     {\mathcal{M}}_{n}~=~  \frac{\langle st \rangle^4\,\d^{(4)}\left(\sum_{i=1}^n\,p_i\right)}{\langle 12\rangle\langle 23 \rangle\cdots \langle n1 \rangle }~~.
\end{align}

Mapping these amplitudes to the celestial sphere requires the following parametrization of momenta
\begin{equation}
    \begin{split}
        \label{eq:momdef}
    p_i^\mu ~&=~ \epsilon_i\, \omega_i\, q^\mu(z_i,\,\bar{z}_i)  \qquad i = 1,\, \dots, \,n\\
     ~&=~\epsilon_i\, \omega_i\, (1+z_i\bar{z}_i,\, (z_i + \bar{z}_i),\, -i(z_i - \bar{z}_i),\ 1-z_i \bar{z}_i )~~,
    \end{split}
\end{equation}
with $\epsilon_i = 1, -1$ for outgoing and incoming particles respectively. The spinor helicity variables, the angle and square brackets can all be written in terms of $z_i, \bar{z}_i, \omega_i$ as
\begin{equation}
    \begin{split}
        \lambda_i^{\a} ~=&~ \epsilon_i\,\sqrt{2\omega_i}\,\begin{pmatrix}
        1\\
        z_i
        \end{pmatrix} ~~,~~
        \lambda_{i,\a} ~=~ \epsilon_i\,\sqrt{2\omega_i}\,\begin{pmatrix}
        -z_i \\
        1
        \end{pmatrix} ~~,\\
        \Tilde{\lambda}_{i,\dot\a} ~=&~  \sqrt{2\omega_i}\,\begin{pmatrix}
        -\Bar{z}_i\\
        1
        \end{pmatrix} ~~,~~
        \Tilde{\lambda}_{i}^{\dot\a} ~=~  \sqrt{2\omega_i}\,\begin{pmatrix}
        1\\
        \Bar{z}_i
        \end{pmatrix} ~~,~~
    \end{split}
     \label{eq:spinors}
\end{equation}
and
\begin{equation}
    \langle i\,j \rangle ~=~ -2\epsilon_i \epsilon_j \sqrt{\omega_i \omega_j}\, z_{ij}~~,~~ \lbrack i\,j \rbrack ~=~ 2\sqrt{\omega_i \omega_j}\, \bar{z}_{ij}~~.
\end{equation}
The celestial superamplitude $\tilde{\mAn}$ is the Mellin transform of the superamplitude w.r.t. to $\omega_i$, 
\begin{align}
    \label{eq:celestialamp}
    \tilde{\mAn}(J_i,\Delta_i,z_i,\Bar{z}_i) ~=~ \int\,\left[ \prod_{i=1}^n \frac{d\omega_i}{\omega_i} \, \omega_i^{\Delta_i} \right]\mAn(h_i,\omega_i,z_i,\Bar{z}_i)~~,
\end{align}
with similar definitions for $\tilde{\calA}_{n,k}$. For more details, see  \cite{Brandhuber:2021nez, Jiang:2021xzy}. 
Here $\Delta_i = 1 + i\lambda_i$ is the conformal dimension and the 2D spin $J_i$ is identified as the 4D helicity of particle $i$, $h_i$. 

In this paper, we will focus only on the celestial MHV amplitudes for which we can use the expression in (\ref{eq:superamplitude}) and (\ref{eq:n-gluonamp}) to write
\begin{equation}
    \begin{split}
        \label{eq:MHVsupercelestial}
    \tilde{\calA}_{n,\,0} ~&=~ \int \left[\prod_{i=1}^n \frac{d\omega_i}{\omega_i}\, \omega_i^{\Delta_i}\right]  \frac{\prod_{A=1}^4\sum_{i,j=1}^n\sqrt{\omega_i \, \omega_j} \,\epsilon_i \, \epsilon_j \, z_{ij} \eta_i^A\eta_j^A}{(-2)^n\, \omega_1 \, \dots \omega_n \, z_{12}\, \dots \, z_{n1} } \delta^{(4)}\left(\sum_{n=1}^n \epsilon_i\,\omega_i \, q_i \right) \\
    ~&=~ \mathcal{H}_{st} \,\int \left[\prod_{i=1}^n \frac{d\omega_i}{\omega_i}\, \omega_i^{\Delta_i}\right] \,\mathcal{M}_n ~=~ \mathcal{H}_{st} \,\,\Tilde{\mathcal{M}}_n~~,
    \end{split}
\end{equation}
with 
\begin{align}
    \label{eq:H12}
    \mathcal{H}_{st} = \frac{1}{z_{st}^4 \,2^4}\prod_{A=1}^4\sum_{i,j=1}^n \left(\epsilon_i \, \epsilon_j \, z_{ij}\right) \eta_i^A\eta_j^A e^{\frac{1}{2}\left(\frac{\partial}{\partial \, \Delta_i} +\frac{\partial}{\partial \, \Delta_j} -   \frac{\partial}{\partial \, \Delta_s} -  \frac{\partial}{\partial \, \Delta_t} \right) } ~~.
\end{align}
$\tilde{\mathcal{M}}_n$ has already been computed in \cite{Schreiber_2018}. We will revisit this in Section~\ref{sec:hyperpde} in greater detail. 

\section{Dual superconformal symmetry}
\label{sec:celestialdc}
\subsection{Dual superconformal symmetry of momentum space amplitudes}
\label{sec:reviewdc}
The remarkable simplicity of scattering amplitudes in $\mathcal{N} = 4$ Yang-Mills theory and our ability to compute them stem partly from the symmetries of the theory. It is now well known that in addition to the conventional superconformal symmetry $PSU(2,2|4)$, tree-level amplitudes of the theory are invariant under dual superconformal symmetry \cite{Drummond:2008vq, Drummond:2009fd} which
can be easily seen if we define dual momentum and supermomentum variables
\begin{align}
    \label{eq:dual momenta}
    &(p_i)_{\alpha\, \dot{\alpha}} ~=~ -\lambda_{i\,\alpha} \, \tilde{\lambda}_{i\,\dot{\alpha}} ~:=~ \left( x_i - x_{i+1} \right)_{\alpha\, \dot{\alpha}}  \qquad i = 1\, \dots, \, n~~,\\
    &q_{i\,\alpha}^A ~=~ \lambda_{i\,\alpha} \eta^A_i ~:=~ \left( \theta_i - \theta_{i+1}\right)_{\alpha}^A \qquad i = 1\, \dots, \,  n ~~.
\end{align}

Dual superconformal symmetry is ordinary superconformal symmetry in variables $(x_i, \theta_i)$. The amplitudes are covariant under this symmetry and the generators do not annihilate them. However, we can modify these generators such that they annihilate the amplitudes and express them in terms of $\lambda_i, \, \tilde{\lambda}_i$, see \cite{Drummond:2008vq, Drummond:2009fd}
for more details. All the generators except $K^{\a \dot{\a}}$ and $S_{\a}^A$ either act trivially on the amplitude or are equal to one of their conformal counterparts. Hence, we only present expressions for $K^{\a \dot{\a}}$ and  $S_{\a}^A$ here. 

Let us first rewrite the expression for the generators  $K^{\a \dot{\a}}$ given in \cite{Drummond:2009fd} in a more compact form 
\begin{equation}
\label{eq:dualK}
    \begin{aligned}
        \calK^{\a \dot{\a}} &= -\sum\limits_{i=1}^n \left\lbrack \sum_{j=1}^{i-1} \lambda_j^{\b} \tilde{\lambda}_j^{\dot{\a}} \lambda_i^{\a} \frac{\p}{\p \lambda_i^{\b}} + \sum\limits_{j=1}^i \lambda_j^{\a} \tilde{\lambda}_j^{\dot{\b}} \tilde{\lambda}_i^{\dot{\a}} \frac{\p}{\p \tilde{\lambda}_i^{\dot{\b}}} + \sum_{j=1}^i \tilde{\lambda}_i^{\dot{\a}} \lambda_j^{\a}\eta_j^A \frac{\p}{\p \eta_i^A} + \sum_{j=1}^{i-1} \lambda_j^{\a} \tilde{\lambda}_j^{\dot{\a}} \right\rbrack\\
      &= -\sum_{i<j} \left( \tilde{\lambda}_i^{\dot{\a}} \lambda_j^{\a} D_{j,i} + \lambda_i^{\a} \tilde{\lambda}_i ^{\dot{\a}} \right)~~,
    \end{aligned}
\end{equation}
where we have made use of momentum conservation and also introduced the operator 
\begin{equation}\label{equ:defdij}
    D_{i,j} ~=~ \lambda_j^{\a}\,\frac{\pa}{\pa \lambda_i^{\a}} ~-~ \Tilde{\lambda}_{i,\Dot \b}\frac{\pa}{\pa \Tilde{\lambda}_{j,\Dot \b}} ~-~ \sum_A\,\eta_i^A\frac{\pa}{\pa \eta_j^A}~~,
\end{equation}
which will play an additional role in Section \ref{sec:diffeqs}.

Similar manipulations on $ \calS_{\a}^A$ lead to 
\begin{equation}
\label{eq:dualS}
    \begin{aligned}
        \calS_{\a}^A &= -\sum_{i=1}^n \left\lbrack \sum_{j=1}^{i-1} \lambda_j^{\b} \eta_j^A \lambda_{i,\a} \frac{\p}{\p \lambda_i^{\b}} + \sum_{j=1}^{i} \lambda_{j,\a}\tilde{\lambda}_j^{\dot{\b}} \eta_i^A \frac{\p}{\p \tilde{\lambda}_i^{\dot{\b}}} - \sum_{j=1}^i \lambda_{j,\a}\eta_{j}^B \eta_i^A \frac{\p}{\p \eta_i^B} + \sum_{j=1}^{i-1} \lambda_{j,\a}\eta_j^A \right\rbrack \\
        &=-\sum_{i<j} \left( \lambda_{j,\a} \eta_i^A D_{j,i} + \lambda_{i,\a} \eta_{i}^A \right)~~.
    \end{aligned}
\end{equation} 

\subsection{Dual superconformal symmetry of celestial amplitudes}
\label{sec:celestialdc2}
Symmetries have played a pivotal role in determining scattering amplitudes in $\mathcal{N}=4$ Yang-Mills. It is conceivable that understanding these symmetries on the celestial sphere will lead to some insight about the putative celestial conformal field theory governing these amplitudes. 
Several results have already been obtained along these lines for Poincar{\'e} \cite{Law:2019glh}, conformal and superconformal symmetries \cite{Stieberger:2018onx, Brandhuber:2021nez, Jiang:2021xzy, Fotopoulos:2020bqj}. Here we take the first steps towards an understanding of the implications of celestial dual superconformal symmetry  by obtaining the form of the generators on the celestial sphere. \\

Let $\calO$ be an operator acting on the amplitude. Then, the corresponding operator $\tilde{\calO}$, which acts on the celestial amplitude is defined by 
\begin{equation} 
    \label{eq:celestialop}
    \begin{aligned}
        &\tilde{\calO} \tilde{\mAn} := \int \left(\prod_{i=1}^n \, \frac{d\omega_i}{\omega_i} \, \omega_i^{\Delta_i} \right)\, \calO \mAn~~.\\
    \end{aligned}
\end{equation}
Then the operator $D_{i,j}$ becomes
\begin{equation}
    \tilde{D}_{i,j} ~=~-\,\epsilon_i\epsilon_j\,e^{\frac{\p}{2\p \Delta_j} - \frac{\p}{2 \p \Delta_i}} \left(\,\Delta_i + J_i + z_{ij}\frac{\pa}{\pa z_i} \,\right) + e^{\frac{\p}{2\p \Delta_i} - \frac{\p}{2 \p \Delta_j}} \left(\,\Delta_j - J_j + \Bar{z}_{ji}\frac{\pa}{\pa \Bar{z}_j} \,\right)- \sum_A\,\eta_i^A\frac{\pa}{\pa \eta_j^A}~~.
\end{equation}
%where $\Delta_i$ and $J_i$ DEFINE HERE OR NEAR 2.9? 

We can use this in (\ref{eq:dualK}) and (\ref{eq:dualS}) to work out the all the components of $\tilde{\calK}^{\alpha \dot{\alpha}}$ and ${\tilde \calS}_\a^A$ on the celestial sphere

\begin{equation}
    \begin{aligned}
        \tilde{\calK}^{\a \dot{\a}} =& \sum\limits_{i<j} \left\{
        \begin{pmatrix}
            1 & \bar{z}_i \\
            z_j & \bar{z}_i z_j\\
        \end{pmatrix}
        \left\lbrack 2\epsilon_i e^{\frac{\pa}{\pa \Delta_i}} \left( \Delta_j + J_j - z_{ij}\frac{\pa}{\pa z_j} \right) + 2 \epsilon_j e^{\frac{\pa}{2\pa \Delta_i} + \frac{\pa}{2\pa \Delta_j}} \sum_A\, \eta_j^A \frac{\pa}{\pa \eta_i^A} \right. \right.\\
        & \qquad\qquad\qquad\qquad\left. \left. - 2\epsilon_j e^{\frac{\pa}{\pa \Delta_j}}\left( \Delta_i - J_i + \bar{z}_{ij} \frac{\pa}{\pa \bar{z}_i} \right) \right\rbrack - 
        2\epsilon_i e^{\frac{\pa}{\pa \Delta_i}}
        \left(
        \begin{aligned}
            & 1 && \bar{z}_i \\
            & z_i && z_i \bar{z}_{i}\\
        \end{aligned}\right) \right\} ~~,
    \end{aligned}
\end{equation}

\begin{equation}
    \begin{aligned}
        {\tilde \calS}_\a^A = & \sqrt{2}\sum\limits_{i<j} \left\{
        \begin{pmatrix}
        -z_j \\
        1
        \end{pmatrix}
        \left\lbrack \epsilon_i \eta_i^A e^{\frac{\pa}{2\pa \Delta_i}} \left( \Delta_j + J_j - z_{ij}\frac{\pa}{\pa z_j} \right) + \epsilon_j e^{ \frac{\pa}{2\pa \Delta_j}} \eta_i^A \sum_B\,\eta_j^B \frac{\pa}{\pa \eta_i^B} \right. \right.\\
        & \qquad\qquad\qquad\left. \left. - \epsilon_j \eta_i^A e^{\frac{\pa}{\pa \Delta_j} - \frac{\pa}{2\pa \Delta_i}}\left( \Delta_i - J_i + \bar{z}_{ij} \frac{\pa}{\pa \bar{z}_i} \right) \right\rbrack - 
        \epsilon_i \eta_i^A e^{\frac{\pa}{2\pa \Delta_i}}
        \begin{pmatrix}
        -z_i \\
        1
        \end{pmatrix} \right\}~~.
    \end{aligned}
\end{equation}

We have explicitly checked that these generators annihilate the all tree-level celestial superamplitudes. 
Note that while the results of the next two sections will be specific to MHV amplitudes, the results of this section
hold for all helicity sectors.

\section{Differential equations for celestial MHV amplitudes\label{sec:diffeqs}}
\subsection{Generalized Banerjee-Ghosh equations}
\label{sec:BGeqs}

In \cite{Banerjee:2020vnt}, Banerjee and Ghosh derived a set of $n-2$ differential equations satisfied by celestial $n$-point MHV gluon amplitudes at 
tree-level. We refer to these as BG equations. While these equations are satisfied by the full amplitude, it is easier to work with the colour ordered amplitudes. We will use the equation derived in \cite{Banerjee:2020vnt} to derive equations for colour ordered MHV amplitudes. Transforming to momentum space, we find that these equations encode the properties of the amplitude under an infinitesimal BCFW shift. We will then derive generalizations of these equations for MHV amplitudes.

\subsubsection{Colour stripped amplitudes}
\label{sec:BG colour stripped}
%In this Appendix, we extract the differential equations satisfied by colour ordered MHV amplitudes (on the celestial sphere) from the one derived in \cite{Banerjee:2020vnt}. 
It is well known that the tree-level $n$-gluon amplitude, $M_n$ admits a colour decomposition
\begin{align}
    \label{eq:colourdecomp}
    M_n = \sum_{\pi\in S_{n-1}}\,{\cal M}_n[1,\pi(2),\pi(3),\dots,\pi(n)]\,{\rm Tr}[T^{a_1}T^{a_{\pi(2)}}\cdots T^{a_{\pi(n)}}]~~,
\end{align}
where $\pi$ is a permutation of $\{1,\,2,\,\cdots,\,n\}$ with the requirement  $\pi(1)=1$. This decomposition trivially carries over to celestial amplitude,
\begin{equation}
    \tilde{M}_n ~:=~ \an{\calO_{\Delta_1}^{a_1} \dots \calO_{\Delta_n}^{a_n}} = \sum_{\pi\in S_{n-1}}\,\tilde{{\cal M}}_n[1,\pi(2),\pi(3),\dots,\pi(n)]\,{\rm Tr}[T^{a_1}T^{a_{\pi(2)}}\cdots T^{a_{\pi(n)}}]~~.
\end{equation}
Here $\tilde{M}_n$ and $\tilde{\cal M}_n$ are the Mellin transforms of $M_n$ and ${\cal M}_n$ respectively. All undefined  symbols used for the remainder of this subsection will have the same meanings as in \cite{Banerjee:2020vnt}. There the authors derived a differential equation for $\tilde{M}_n$. To do this, they imposed the constraints arising from the current algebra of the subleading soft modes of a gluon primary on the subleading singularities of the OPE between a gluon primary and a subleading soft mode. This leads to the identification of a null state, which when inserted into a correlation function of $n$ gluon primaries, leads to a differential equation. We follow this route here albeit with minor modifications to arrive at differential equations for the colour ordered amplitude $\tilde{\cal M}_n \left(1, \dots n\right)$. 

We begin with equation (6.5) from \cite{Banerjee:2020vnt}, which after  making the replacement $1 \to i$ is  
\begin{align}
    \label{eq:BGstart}\
    \left[ \delta^{a_i \,x} J^a_{-1} \, + \, i\, f^{a\, a_i\,x} L_{-1} P^{-1}_i - \left( \delta^{a\,x} \delta^{a_i\,y} - (\Delta_i - 1) \delta^{a\, y}\delta^{a_i\,x} \right) j_{-1}^y \, P_i^{-1}  \right] \, \calO_{\Delta_i, +}^x = 0~~.
\end{align} 
Shifting $\Delta_i \to \Delta_i + 1$, using $\calO_{\Delta_i+1, +}^x = P_i \calO_{\Delta_i, +}^x$ and inserting this into the MHV amplitude yields  
\begin{equation}
    \begin{split}
       \label{eq:BGMHV}
    &\left[ \delta^{a_i \,x} J^a_{-1}(i) \,P(i)  ~-~ \left( \delta^{a\,x} \delta^{a_i\,y} - \Delta_i  \delta^{a\, y}\delta^{a_i\,x} \right) j_{-1}^y(i)   \right]\,\an{\calO_{\Delta_1}^{a_1} \dots \calO_{\Delta_i}^{x} \dots \calO_{\Delta_n}^{a_n}} \\ 
    &\hspace{6cm}~+~  T^a(i) L_{-1}(i)\an{\calO_{\Delta_1}^{a_1} \dots \calO_{\Delta_i}^{a_i} \dots \calO_{\Delta_n}^{a_n}} ~=~ 0~~.
    \end{split}
\end{equation}
Performing the appropriate OPEs, this turns into a differential equation for $\tilde{M}_n$
\begin{equation}
    \begin{split}
      &~T_i^a \, \frac{\pa}{\pa z_i} \,\an{\calO_{\Delta_1}^{a_1} \dots \calO_{\Delta_i}^{a_i} \dots \calO_{\Delta_n}^{a_n}} ~-~ \sum_{j\ne i}\epsilon_i\epsilon_j\,\frac{\Delta_j - J_j-1+\Bar{z}_{ji}\Bar{\pa}_j}{z_{ji}}\,T_j^a\an{\calO_{\Delta_1}^{a_1} \dots \calO_{\Delta_{j-1}}^{a_{j}} \dots \calO_{\Delta_{i+1}}^{a_{i}} \dots \calO_{\Delta_n}^{a_n}}\\
       &~-~\sum_{j\ne i}\,\frac{T_j^{a_i}}{z_{ji}}\,\an{\calO_{\Delta_1}^{a_1} \dots \calO_{\Delta_i}^{a} \dots \calO_{\Delta_n}^{a_n}}~+~ \Delta_i\,\sum_{j\ne i}\,\frac{T_j^{a}}{z_{ji}}\,\an{\calO_{\Delta_1}^{a_1} \dots \calO_{\Delta_i}^{a_i} \dots \calO_{\Delta_n}^{a_n}} = 0~~.
    \end{split}
    \label{equ:original-PDE}
\end{equation}
This is true for each positive helicity gluon $i$. Thus we have a set of $(n-2)$ partial differential equations satisfied by $\tilde{M}_n$. We can now extract the equations satisfied by the colour ordered amplitudes. For the sake of concreteness, we will derive equations for the colour ordered celestial amplitude $\tilde{\cal M}_n \left( 1\, 2\, \dots n\right)$. The relevant contribution from the first term in (\ref{equ:original-PDE}) is
\begin{align}
    \begin{split}
       &~\left(\frac{\partial}{\partial \, z_i }  \tilde{\cal M}_n \left(1 \, \dots \, n\right) \right) \,  T_i^a\,{\rm Tr}[T^{a_1}\cdots T^{a_{i-1}}T^xT^{a_{i+1}}\cdots]\\
       ~=&~ \left(\frac{\partial}{\partial \, z_i }  \tilde{\cal M}_n \left(1 \, \dots \, n\right) \right) \, i\,f^{aa_i x}\,{\rm Tr}[T^{a_1}\cdots T^{a_{i-1}}T^xT^{a_{i+1}}\cdots] \\
        ~=&~\left(\frac{\partial}{\partial \, z_i }  \tilde{\cal M}_n \left(1 \, \dots \, n\right) \right) \,\left( {\rm Tr}[T^{a_1}\cdots T^{a_{i-1}}\textcolor{red}{T^a}T^{a_i}T^{a_{i+1}}\cdots]~-~{\rm Tr}[T^{a_1}\cdots T^{a_{i-1}}T^{a_i}\textcolor{red}{T^a}T^{a_{i+1}}\cdots]\right) ~~.
    \end{split}
\end{align} 
Looking at terms with the prefactor of ${\rm Tr}[T^{a_1}\cdots T^{a_{i-1}}\textcolor{red}{T^a}T^{a_i}T^{a_{i+1}}\cdots]$, we see that the next three terms in eq (\ref{equ:original-PDE}) contribute only when $j = i \pm 1$ yielding
\begin{equation}
    \begin{split}
       \label{eq:colourstripped1}
& \left(\pa_i ~-~\frac{\Delta_i}{z_{i-1,i}} ~-~ \frac{1}{z_{i+1,i}}\right)\tilde{{\cal M}}_n(1,\cdots,n)\\
 &\hspace{.4cm} ~+~ \left(\epsilon_i\epsilon_{i-1}\,\frac{\Delta_{i-1}-J_{i-1}-1+\Bar{z}_{i-1,i}\Bar{\pa}_{i-1}}{z_{i-1,i}} e^{\frac{\pa}{\p \Delta_i} - \frac{\pa}{\pa \Delta_{i-1}}}\right)\tilde{{\cal M}}_n(1,\cdots,n) ~=~ 0 ~~,
    \end{split}
\end{equation}
and another equation with $i-1 \leftrightarrow i+1$ if we look at terms with the prefactor of ${\rm Tr}[T^{a_1}\cdots T^{a_{i-1}}T^{a_i}\textcolor{red}{T^a}T^{a_{i+1}}\cdots]$. The terms with the prefactor of\\ ${\rm Tr}[T^{a_1}\cdots T^{a_{i-1}}T^{a_i}T^{a_{i+1}}\cdots T^{a_{j-1}}\textcolor{red}{T^a}T^{a_j}T^{a_{j+1}}\cdots]$, in which the insertion $T^a$ isn't adjacent to $T^{a_i}$ imply  

\begin{equation}
    \begin{split}
      &\left\lbrace -\,\epsilon_i\epsilon_{j}\,\frac{\Delta_{j} - J_{j} -1+\Bar{z}_{j,i}\Bar{\pa}_{j}}{z_{j,i}}\,e^{\frac{\pa}{\pa \Delta_i} - \frac{\pa}{\pa \Delta_j}}
       ~+~ \epsilon_i\epsilon_{j-1}\,\frac{\Delta_{j-1} - J_{j-1} -1+\Bar{z}_{j-1,i}\Bar{\pa}_{j-1}}{z_{j-1,i}}\,e^{\frac{\pa}{\pa \Delta_i} - \frac{\pa}{\pa \Delta_{j-1}}}\right.\\
       &\hspace{.2cm}~+~\left.\left(\,\frac{\Delta_i}{z_{j,i}}~-~\frac{\Delta_i}{z_{j-1,i}}\,\right)\right\rbrace\,{\tilde{\cal M}}_n[1,\cdots,i-1,i,i+1,\dots,j-1,j,\cdots]\\
       &\hspace{1cm}~+~  \left(\,\frac{1}{z_{i-1,i}}~-~\frac{1}{z_{i+1,i}}\,\right)\,{\tilde{\cal M}}_n[1,\cdots,i-1,i+1,\dots,j-1,i,j,\cdots] = 0~~.
    \end{split}\label{eq:colourstripped3}
\end{equation}
Plugging in the form of the MHV celestial amplitude (\ref{eq:MHVmellin}), it is easy to see that (\ref{eq:colourstripped3}) is satisfied if (\ref{eq:colourstripped1}) is true. Thus, we will take (\ref{eq:colourstripped1}) to be the set of independent equations satisfied by the colour stripped amplitude. 

\subsubsection{Momentum space origin}

We can recast (\ref{eq:colourstripped1}) as an equation acting on the momentum space amplitude ${\cal M}_{n}$ to get 
\begin{align}
    \label{eq:momspaceBG}
 \left(   \lambda_{i-1}^{\alpha} \frac{\partial}{\partial \lambda_i^{\alpha}} -     \tilde{\lambda}_{i}^{\dot{\alpha}} \frac{\partial}{\partial \tilde{\lambda}_{i-1}^{\dot{\alpha}}} \right) {\cal M}_{n}= \frac{\an{i-1\,i+1}}{\an{i+1\,i}} {\cal M}_{n} ~~,
\end{align}
where $i$ is a positive helicity gluon. We will now demonstrate that this equation follows simply from the properties of ${\cal M}_{n}$ under a BCFW shift and that we can generalize this equation to superamplitudes.   
%%%%%%%%%%%%%%%%%%%%% non-supersymmetric %%%%%%%%%%%%%%%%%%%%%%%%%

\label{sec:BGnonsusy}
Consider the $n$-point MHV amplitude with negative helicity gluons $s$ and $t$
\begin{equation}
  {\cal M}_{n}[123\dots n]~=~  \frac{\langle st \rangle^4\,\d^{(4)}\left(\sum_{i=1}^n\,p_i\right)}{\langle 12\rangle\langle 23 \rangle\cdots \langle n1 \rangle }~~,
  \label{equ:n-pt-MHV}
\end{equation}
and the BCFW shift
\begin{align}
    \label{eq:BCFW shift}
    \lambda_i ~\to~ \hat{\lambda}_i ~=~  \lambda_i \, + \, z\, \lambda_j ~~,~~ \tilde{\lambda}_j ~\to~ \hat{\tilde{\lambda}}_j  ~=~ \tilde{\lambda}_j \, - \,z \, \tilde{\lambda}_i~~. 
\end{align}
For infinitesimal $z$, this shift is implemented on (\ref{equ:n-pt-MHV}) by the operator
\begin{equation}
    D_{i,j} ~=~ \lambda_j^{\a}\,\frac{\pa}{\pa \lambda_i^{\a}} ~-~ \Tilde{\lambda}_{i,\Dot \b}\frac{\pa}{\pa \Tilde{\lambda}_{j,\Dot \b}}  ~~~.
\end{equation}
Note that this coincides with the LHS of (\ref{eq:momspaceBG}) when $j=i-1$ and $i$ corresponds to a negative helicity gluon. The RHS depends on the helicities of $i, j$ and we tabulate the cases below.
\begin{enumerate}
    \item When $(J_i,\,J_j) = (+,\pm),\,(-,\,-)$,
    \begin{equation}
       \begin{split}
            \hat{\cal M}_{n}[\dots \hat{i} \dots] ~=&~ {\cal M}_{n}\,\frac{1}{1+z\,\frac{\langle i-1,j \rangle}{\langle i-1,i \rangle}}\,\frac{1}{1+z\,\frac{\langle i+1,j \rangle}{\langle i+1,i \rangle}} \\
            ~=&~ {\cal M}_{n}\,\left(1-z\,\frac{\langle i-1,j \rangle}{\langle i-1,i \rangle} -z\, \frac{\langle i+1,j \rangle}{\langle i+1,i \rangle} \right) ~+~ {\cal O}(z^2) ~~.
       \end{split}
    \end{equation}
    %where the notation $i_l = \pi(\pi^{-1}(i)-1)$ meaning the particle $i_l$ sitting on the left of particle $i$ in this ordering, and $i_r = \pi(\pi^{-1}(i)+1)$ meaning the particle $i_r$ sitting on the right of particle $i$ in this ordering.
    \item When $(J_i,\,J_j) = (-,\,+)$, consider $i=s$,
    \begin{equation}
       \begin{split}
            \hat{\cal M}_{n}[\dots \hat{i} \dots] ~=&~ \frac{(\langle it \rangle+z\,\langle jt \rangle)^4}{\langle 12\rangle\langle 23\rangle\cdots \langle n1 \rangle }\,\frac{1}{1+z\,\frac{\langle i-1,j \rangle}{\langle i-1,i \rangle}}\,\frac{1}{1+z\,\frac{\langle i+1,j \rangle}{\langle i+1,i \rangle}} \\
            ~=&~ {\cal M}_{n}\,\left(1-z\,\frac{\langle i-1,j \rangle}{\langle i-1,i \rangle} -z\, \frac{\langle i+1,j \rangle}{\langle i+1,i \rangle}+ 4z\,\frac{\langle j,t \rangle}{\langle i,t \rangle} \right) ~+~ {\cal O}(z^2) ~~.
       \end{split}
    \end{equation}
\end{enumerate}
Both of these cases can be compactly summarized as 
\begin{equation}
    D_{i,j}\, {\cal M}_{n} ~=~ {\cal M}_{n}\,\left(-\,\frac{\langle i-1,j \rangle}{\langle i-1,i \rangle} - \frac{\langle i+1,j \rangle}{\langle i+1,i \rangle} + 4\,\frac{\langle j,t \rangle}{\langle i,t \rangle}\,\d_{i,s}+ 4\,\frac{\langle j,s \rangle}{\langle i,s \rangle}\,\d_{i,t} \right)~~,
    \label{equ:BCFW-pde-MHV}
\end{equation}
which is a generalization of (\ref{eq:momspaceBG}). \\

%%%%%%%%%%%%%%%%%%%%% reduce to BG equation %%%%%%%%%%%%%%%%%%%%%%%%%

We can map this to the celestial sphere. Taking $J_i = +$, we get
\begin{equation}
    \begin{split}
        &\left[ -\left(\Delta_i +z_{ij}\frac{\pa}{\pa z_i} \right) ~+~ \frac{z_{i-1,j}}{z_{i-1,i}}  ~+~ \frac{z_{i+1,j}}{z_{i+1,i}} ~-~ 1 \right]\,\Tilde{\cal M}_{n}  \\
        & \quad\quad ~+~ \epsilon_i\epsilon_j\left(\Delta_j - J_j-1+\Bar{z}_{ji}\frac{\pa}{\pa \Bar{z}_j} \right)\,e^{\frac{\partial}{\partial \Delta_i} - \frac{\partial}{\partial \Delta_j}}\Tilde{\cal M}_{n} ~=~ 0~~,
    \end{split}
    \label{equ:general-BG}
\end{equation}
which generalizes (\ref{eq:colourstripped1}).

\smallskip

\subsubsection{Supersymmetric case}

%\smallskip

\label{sec:BGsusy}
The discussion above readily generalizes to MHV superamplitudes. The super-BCFW shift  
\begin{equation}
    \lambda_i ~\rightarrow~ \lambda_i~+~z\,\lambda_j ~~,~~ \Tilde{\lambda}_j ~\rightarrow~ \Tilde{\lambda}_j ~-~z\,\Tilde{\lambda}_i~~,~~\eta^A_j ~\rightarrow~\eta^A_j ~-~ z\,\eta^A_i
    \label{equ:super-BCFW-shift}
\end{equation}
is generated by
\begin{equation}\label{equ:defdij-super}
    D_{i,j} ~=~ \lambda_j^{\a}\,\frac{\pa}{\pa \lambda_i^{\a}} ~-~ \Tilde{\lambda}_{i,\Dot \b}\frac{\pa}{\pa \Tilde{\lambda}_{j,\Dot \b}} ~-~ \sum_A\,\eta_i^A\frac{\pa}{\pa \eta_j^A} ~~~.
\end{equation}
Recall that
\begin{equation}
    {\cal A}_{n,0} ~=~ \frac{\frac{1}{2^4}\prod_{A=1}^4\sum_{i,j=1}^n\langle ij \rangle \eta_i^A\eta_j^A}{\langle 12\rangle\langle 23 \rangle\cdots \langle n1 \rangle } \delta^{(4)} \left( \sum_{i=1}^n p_i \right)~~.
    \label{equ:superamplitude}
\end{equation}
Under the super-BCFW shift (\ref{equ:super-BCFW-shift}), the numerator of (\ref{equ:superamplitude}) is invariant and we have
\begin{equation}
    D_{i,j}\, {\cal A}_{n,0} ~=~ -\,{\cal A}_{n,0}\,\left(\frac{\langle i-1,j \rangle}{\langle i-1,i \rangle} + \frac{\langle i+1,j \rangle}{\langle i+1,i \rangle} \right)~~.
    \label{equ:super-BCFW-pde-MHV}
\end{equation}
Note that the equation above is valid for arbitrary $i$ and $j$. \\

Transforming (\ref{equ:super-BCFW-pde-MHV}) to the celestial sphere we get 
\begin{equation}
    \begin{split}
        &\left[ -\left(\Delta_i + J_i+z_{ij}\frac{\pa}{\pa z_i} \right) ~+~ \frac{z_{i-1,j}}{z_{i-1,i}}  ~+~ \frac{z_{i+1,j}}{z_{i+1,i}}   \right]\,\Tilde{\cal A}_{n,0}  \\
        & \quad\quad~-~ \epsilon_i\epsilon_j\,\left(\sum_A\,\eta_i^A\frac{\pa}{\pa \eta_j^A}\right)e^{\frac{1}{2}\frac{\partial}{\partial \Delta_i} - \frac{1}{2}\frac{\partial}{\partial \Delta_j}} \,\Tilde{\cal A}_{n,0}\\
        & \quad\quad ~+~ \epsilon_i\epsilon_j\left(\Delta_j - J_j-1+\Bar{z}_{ji}\frac{\pa}{\pa \Bar{z}_j} \right)\,e^{\frac{\partial}{\partial \Delta_i} - \frac{\partial}{\partial \Delta_j}}\Tilde{\cal A}_{n,0} ~=~ 0~~,
    \end{split}
    \label{equ:general-BG-super}
\end{equation}
which is a generalization of (\ref{eq:colourstripped1}) to celestial MHV superamplitudes.

\subsection{Hypergeometric equations and their momentum space origin}
\label{sec:hyperpde}
The tree-level celestial MHV gluon $n$-point amplitudes have been computed in \cite{Schreiber_2018} and have been found to be Aomoto-Gelfand hypergeometric functions. It is well known that these functions satisfy a set of differential equations. We refer the reader to Appendix~\ref{sec:AGfunc} for more details. In this section, we will motivate these differential equations from properties of amplitudes. The celestial tree-level MHV $n$-point amplitude is given by

\begin{equation}
    \begin{split}
        \label{eq:MHVmellin}
    \tilde{\mathcal{M}}_{n} ~=&~ 2\pi\, \frac{z_{st}^4}{z_{12}z_{23}\dots z_{n1}} \frac{\delta\left(\sum_i (\Delta_i - 1)\right)}{2^{n-4}} \frac{1}{U} \prod_{a,b} \mathbf{1} \left( x_{a,b} >0\right)\\
    &~\times\int \left(\prod_{c=1}^{n-4} \, \D u_c\right) \,  \delta\left(1-\sum_{d=1}^{n-4}u_d\right) \, \left(\prod_{a=1}^{n-4} u_a^{\Delta_a-J_a-1}\right) \prod_{b=n-3}^n \left(\sum_{a=1}^{n-4} x_{a,b}u_a\right)^{\Delta_b-J_b-1}\\
    &~:= \mathcal{N} \,\delta\left(\sum_i (\Delta_i - 1)\right) F \left( x_{a,\,b}, \Delta_i\right)~~,
    \end{split}
\end{equation}
where 
\begin{equation}\label{eq:defxab}
    \begin{aligned}
        &U=\det\{q_{n-3},q_{n-2},q_{n-1},q_n\},\ U_{a,b} = \det\{q_{n-3},q_{n-2},q_{n-1},q_n\}|_{b \to a}~~,\\
        &x_{a,b} = -\frac{ \epsilon_a\, U_{a,b}}{ \epsilon_b\, U }~~,
    \end{aligned}
\end{equation}
 and $F$ is the Aomoto-Gelfand hypergeometric function\cite{aomoto}. This is of the form given in (\ref{eq:Ahypergeometric}) with 
\begin{align}
    \label{eq:MHVahyper}
    Z ~=~ \begin{pmatrix}
    1 & 0 & 0 & \dots & 0 & x_{1,\,n-3} & x_{1,\,n-2} & x_{1,\,n-1 } & x_{1,\,n}\\
    0 & 1 & 0 & \dots & 0 & x_{2,\,n-3} & x_{2,\,n-2} & x_{2,\,n-1} & x_{2,\,n}\\
    \vdots \\
    0 & 0 & 0 & \dots & 1 & x_{n-4,\,n-3} & x_{n-4,\,n-2} & x_{n-4,\,n-1} & x_{n-4,\,n}\\
    \end{pmatrix} ~~,
\end{align}
which corresponds to setting (with $ k = n-4$ and $m = n$)
\begin{equation}
    \begin{split}
         \label{eq:gauge}
& z_{a,b} ~=~ \delta_{a,b} \qquad 1 \leq a ,\, b \leq n-4  ~~,  \\
 & z_{a,b} ~=~ x_{a,b} \qquad 1 \leq a \leq n-4 ~~,~~ n-3 \leq b \leq n~~.
    \end{split}
\end{equation}
The polynomials $\ell_b$ are  
\begin{equation}
    \begin{split}
        &\ell_1 ~=~ u_1,\qquad \dots \ , \qquad \ell_{n-4} ~=~ u_{n-4}~,\\
&\ell_b ~=~ \sum_{a=1}^{n-4} x_{a,b} u_a \,, \qquad b ~=~ {n-3, \dots , n}~~,
    \end{split}
\end{equation}
and the exponents are given by 
\begin{align}
    \label{eq:exponents}
    \alpha_j ~=~ \Delta_j - J_j - 1~~.
\end{align}

Momentum conservation is manifest in (\ref{eq:celestialamp}) due to the presence of the delta function in $\calA_n$. However, in arriving at the expression in (\ref{eq:MHVmellin}), we have used the delta function to eliminate $\omega_{n-3}, \, \dots ,\omega_n$. The statement of momentum conservation can now be written as a differential equation on $F$. Let $\mathbb{P}^\mu$ be the total momentum operator acting on ${\mathcal{M}}_{n}$. The corresponding celestial operator, along the lines of eq (\ref{eq:celestialop}) is
 \begin{align}
     \label{eq:celestialmomop}
     \tilde{\mathbb{P}}^\mu  = \sum_{i=1}^n \epsilon_i\,q_i^\mu e^{\frac{\partial}{\partial \, \Delta_i}} ~~,
 \end{align}
 and we must have
 \begin{align}
     \label{eq:celestialmomcons}
     \sum_{i=1}^n \epsilon_i\,q_i^\mu e^{\frac{\partial}{\partial \, \Delta_i}} \Tilde{\cal M}_{n} = 0 ~~.
 \end{align}
 We can extract four independent conditions by contracting this with the four vectors $v_b^\mu$, $b = \left\lbrace n-3, \, n-2, \, n-1,\, n\right\rbrace$ defined by  
 \begin{equation}
     \begin{split}
         \label{eq:momconsbasis}
     &\epsilon_{n}\,v_{n}^\mu = -\,\epsilon^{ \mu\,\nu\, \rho \, \sigma} q_{n-3\, \nu} q_{n-2\, \rho} q_{n-1\, \s}  \qquad,~~~ \epsilon_{n-3}\,v_{n-3}^\mu =  \epsilon^{\mu\, \nu\, \rho \, \sigma} q_{n-2\, \nu}q_{n-1\, \rho}q_{n\, \s}~~,\\
     &\epsilon_{n-2}\,v_{n-2}^\mu = -\,\epsilon^{\mu\, \nu\, \rho \, \sigma} q_{n-1\, \nu}q_{n\, \rho}q_{n-3\, \s}
 ~~~,~~~  \epsilon_{n-1}\,v_{n-1}^\mu =\epsilon^{\mu\, \nu\, \rho \, \sigma} q_{n\, \nu}q_{n-3\, \rho}q_{n-2\, \s}~~.
     \end{split}
 \end{equation}
 Noting that $v_b^{\mu}\,\epsilon_{a}\, q_{a\mu} = -U x_{a,b}$, we get

\begin{equation}
    \begin{split}
        \label{eq:momconstoHPDE}
  &\left[ - \sum_{a=1}^{n-4} x_{a,b} \, U \, e^{\frac{\partial}{\partial \, \Delta_a}} + U\, e^{\frac{\partial}{\partial \, \Delta_b}} \right]  \,\Tilde{\cal M}_{n} = 0 \\ 
  & \implies  \mathcal{N}\, \delta \left( 1 + \sum_{i=1}^n \left( \Delta_i - 1\right) \right) e^{\frac{\partial}{\partial \Delta_b}} \left(-\sum_{a=1}^{n-4} \frac{1}{\alpha_b} x_{a,b} \frac{\partial}{\partial x_{a,b}} + 1 \right)  F = 0~~,
    \end{split}
\end{equation}
where we have used the fact that $ e^{\frac{\partial}{\partial \Delta_a}} F = e^{\frac{\partial}{\partial \Delta_b}} \frac{1}{\alpha_b} \frac{\partial}{\partial x_{a,b}} F$. This gives us a set of differential equations
 \begin{align}
     \label{eq:HFPDE1}
      \sum_{a=1}^{n-4} x_{a,b} \frac{\partial \, F}{\partial x_{a,b}} = \alpha_b\, F~~,
 \end{align}
 which correspond to eq (\ref{eq:diffscale}).
 
The other set of differential equations eq (\ref{eq:diffglk}) can be derived by noting that the form for $\Tilde{\cal M}_{n}$ given in (\ref{eq:MHVmellin}) corresponds to using the momentum conserving delta function to solve for $\omega_{n-3}, \dots , \omega_n$. A different choice, say solving for $\omega_k, \dots , \omega_{k+3}$ would lead to 

\begin{equation}
   \begin{split}
       \label{eq:MHVmellin2}
    \Tilde{\cal M}_{n} ~=&~2\pi\, \frac{z_{s,t}^4}{z_{12}z_{23}\dots z_{n1}} \frac{\delta\left(\sum_i (\Delta_i - 1)\right)}{2^{n-4}} \frac{1}{U'} \, \int \left[\prod_{c\neq \lbrace k, \, \dots,\, k+3\rbrace} \, \D u_c \right] \,  \delta\left(1-\sum_{d\neq \lbrace k, \, \dots,\, k+3\rbrace}u_d\right) \\
    &~\times~\left[\prod_{b\neq \lbrace k, \, \dots,\, k+3\rbrace} u_b^{\Delta_b-s_b-1}\right] \prod_{b=k}^{k+3} \left(\sum_{a \neq \lbrace k \, \dots k+3 \rbrace} x'_{a,b}u_a\right)^{\Delta_b-s_b-1},
    \end{split}
\end{equation}
with
\begin{equation}
    \begin{aligned}
        &U'=\det\{q_k , q_{k+1}, q_{k+2}, q_{k+3}\}~,\ U'_{a,b} = \det\{q_k , q_{k+1}, q_{k+2}, q_{k+3}\}|_{b \to a}~~,\\
        &x'_{a,b} = -\frac{ \epsilon_a\, U'_{a,b}}{ \epsilon_b\, U' }~~.
    \end{aligned}
\end{equation}
This corresponds to an integral of the form given in (\ref{eq:Ahypergeometric}) but with 
\begin{align}
    \label{eq:MHVahyper2}
    Z ~=~ \begin{pmatrix}
    1 & 0 & \dots & x_{1,\,k} & x_{1,\,k+1} & x_{1,\, k+2} & x_{1,\, k+3} & \dots & 0\\
    0 & 1  & \dots & x_{2,\, k} & x_{2,\, k+1} & x_{2,\,k+2} & x_{2,\,k+3} & \dots & 0\\
    \vdots \\
    0 & 0 & \dots & x_{n-4,\,k} & x_{n-4,\,k+1} & x_{n-4,\,k+2} & x_{n-4,\,k+3} & \dots & 1\\
    \end{pmatrix} ~~.
\end{align}
Both of these are equivalent representations of the celestial amplitude and $Z$ is defined up to $GL(n-4)$ transformations acting on the left. These transformations correspond to a change of solved and unsolved variables. As shown in Appendix~\ref{sec:AGfunc}, this transformation property of $\tilde{\mathcal{M}}_n$ gives rise to the differential equation (\ref{eq:diffglk}).

\begin{align}
    \label{eq:diff1}
     &\sum_{j=1}^n z_{a,b} \frac{\partial \, F}{\partial z_{c,b}} ~=~ -\delta_{a,c} \, F \qquad 1 \, \leq a,\,c \, \leq n-4 ~~.
\end{align}
We need to make sense of the differentials of the gauge fixed $z_{i,j}$. To do this, we first differentiate the function $F$ as in (\ref{eq:Ahypergeometric}) and then set $z_{i,\,j}$ to the corresponding gauge fixed values. In particular, $z_{a,b} \frac{\p F}{\p z_{c,b}} = \delta_{a,b}\, \alpha_a e^{\frac{\pa}{\pa \Delta_c} - \frac{\pa}{\pa \Delta_a}} F$ for $a,\,b,\,c \in \left\lbrace 1, \, \dots n-4\right\rbrace$. We can now write (\ref{eq:diff1}) in terms of the $x_{a,b}$.
\begin{align}
    &\alpha_a F + \sum_{b=n-3}^n x_{a,b} \frac{\partial \, F}{\partial x_{a,b}} ~=~ -F \label{eq:HFPDE21}~~, \\
    &\alpha_a e^{\frac{\partial}{\partial \Delta_c} - \frac{\partial}{\partial \Delta_a}} F + \sum_{b=n-3}^n x_{a,b} \frac{\partial \, F}{\partial \, x_{c,b}} ~=~ 0 \qquad c \neq a ~~. \label{eq:HFPDE22}
\end{align}
It can be shown that (\ref{eq:HFPDE22}) can be derived from the rest and isn't independent.\\

\subsection{Relation between differential equations}
\label{sec:relation}

Finally, we make a comment here about the relationship between the colour stripped BG equations (\ref{eq:colourstripped1}) and the hypergeometric PDEs. Recall that the celestial MHV amplitude 
\begin{align}
    \tilde{\mathcal{M}}_{n} ~=&~  \mathcal{N} \,\delta\left(\sum_i (\Delta_i - 1)\right) F \left( x_{a,\,b}, \Delta_i\right)~~,
\end{align}
where $F$ is a hypergeometric function. To compare the two sets of equations, we rewrite the BG equations as equations for $F$ in terms of the variables $x_{a,\,b}$ in (\ref{eq:defxab}). Without loss of generality, we choose $i=1$. After some manipulation, it can be brought to the form  
\begin{equation}\label{eq:bgmhv33}
    \begin{aligned}
        &\left\lbrack \left( \alpha_1 + 1 + \sum_{b=n-3}^n x_{1,b} \frac{\pa}{\pa x_{1,b}} \right) - \sum_b \frac{\epsilon_2}{\epsilon_1}\left(x_{2,b} + \bar{z}_{1,2} \frac{\pa x_{2,b}}{\pa \bar{z}_2} \right) \left( \frac{\pa}{\pa x_{1,b}} - \frac{\pa}{\pa x_{2,b}}e^{\frac{\pa}{\pa \Delta_1} - \frac{\pa}{\pa \Delta_2}} \right) \right. \\
        &\left.\qquad\qquad\qquad\qquad\qquad\qquad\qquad\qquad - \frac{\epsilon_2}{\epsilon_1} e^{\frac{\pa}{\pa \Delta_1} - \frac{\pa}{\pa \Delta_2}} \left( \alpha_2 + 1 + \sum_b x_{2,b} \frac{\pa}{\pa x_{2,b}} \right) \right\rbrack F = 0~~ ,
    \end{aligned}
\end{equation} 
which shows that the BG equations reduce to combinations of the hypergeometric equations.

\acknowledgments
We are grateful to Shamik Banerjee and Sudip Ghosh for correspondence in the early stages of this work. We thank Jorge Mago, Anders Schreiber and Marcus Spradlin for many stimulating discussions. This work was supported in part by the US Department of Energy under contract {DE}-{SC}0010010 Task A and by Simons Investigator Award \#376208.
The research of Y. Hu is supported in part by the endowment from the Ford Foundation Professorship of Physics and she acknowledges the support of the Brown Theoretical Physics Center.

\appendix

\section{Aomoto-Gelfand hypergeometric functions}
\label{sec:AGfunc}
The Aomoto-Gelfand hypergeometric functions \cite{aomoto} are associated to Grassmannians. In this section, we will make this connection explicit and derive the differential equations associated to them. For a quick introduction, see \cite{Abe} from which most of this section is adopted (with cosmetic changes).\\

Let $Z$ be a $k \times m$ matrix,
\begin{align}
\label{eq:Xmatrix}
    Z = \begin{pmatrix}
    z_{1,1} & z_{1,2} & \dots & z_{1,m}\\
    \vdots\\
    z_{k,1} & z_{l,2} & \dots & z_{k,m}
    \end{pmatrix}~~.
\end{align}
The hypergeometric function associated to this matrix is 
\begin{align}
    \label{eq:Ahypergeometric}
    F(Z) = \int \prod_{b=1}^m \ell_b (u)^{\alpha_b}  \frac{du_1 \dots du_k}{\text{Vol GL(1)}}~~,
\end{align}
where $\ell_b (u) = \sum_{a=1}^k u_a \, z_{a,b}$ and $u= \left( u_1, \dots u_k \right) \in \mathbb{CP}^k$. The polynomials $\ell_1, \, \dots, \, \ell_m$ can be collectively written as 
\begin{align}
    \label{eq:polynomials}
    \ell = \left( \ell_1, \, \dots \ell_m \right) = u\cdot Z~~.
\end{align}
A point in the Grassmannian $G(k,m)$ is defined modulo $GL(k)$ transformations. We would like to understand how the function $F(Z)$ behaves under such transformations. A $GL(k)$ transformation, $G$, acts by left multiplication on the matrix $Z$ as $Z \to G \cdot Z$ and we have,
\begin{align}
   \label{eq:glktranform}
   \ell \to \ell' = u\cdot G \, Z~~.  
\end{align}
We see that the effect on $F(Z)$ is a change of variables from $u\to u'$ leading to 
\begin{align}
\label{eq: glk}
F(G \cdot Z) = \text{det} (G)^{-1} F(Z)~~.
\end{align}
We can also rescale each column separately. This is effected through right multiplication by an $m \times m$ diagonal matrix $S = \text{diag} \left( s_0, \dots s_m \right)$.
\begin{align}
    \label{eq:scalingtransform}
    Z \to Z\cdot S =   \begin{pmatrix}
    s_1 \,z_{1,1} & s_2 \,z_{1,2} & \dots & s_m \, z_{1,m}\\
    \vdots\\
    s_1 \, z_{k,1} & s_2 \, z_{l,2} & \dots & s_m \, z_{k,m}
    \end{pmatrix}~~.
\end{align}
The polynomials transform in a simple manner as $\ell_i \to s_i \, \ell_i$ resulting in 
\begin{align}
    \label{eq:scaling}
    F(Z \cdot S) = F(Z)\,\prod_{b=1}^m s_b^{\alpha_b}~~.
\end{align}
The transformation properties in eqs(\ref{eq: glk}, \ref{eq:scaling}) lead to the following set of differential equations \cite{aomoto}, \cite{Abe} 
\begin{align}
    \label{eq:diffglk}
    &\sum_{b=1}^m z_{a,b} \frac{\partial \, F}{\partial z_{c,b}} = -\delta_{a,c} \, F \qquad 1 \, \leq a,\,c \, \leq k~~,\\
\label{eq:diffscale}
    & \sum_{a=1}^k z_{a,b} \frac{\partial \, F}{\partial \, z_{a,b}} = \alpha_b \, F \qquad 1\, \leq b \, \leq m~~.
\end{align}

\bibliography{main}

\providecommand{\href}[2]{#2}\begingroup\raggedright\begin{thebibliography}{10}

\bibitem{He:2015zea}
T.~He, P.~Mitra, and A.~Strominger, {\it {2D Kac-Moody Symmetry of 4D
  Yang-Mills Theory}},  {\em JHEP} {\bf 10} (2016) 137,
  [\href{http://arxiv.org/abs/1503.02663}{{\tt arXiv:1503.02663}}].

\bibitem{Kapec:2016jld}
D.~Kapec, P.~Mitra, A.-M. Raclariu, and A.~Strominger, {\it {2D Stress Tensor
  for 4D Gravity}},  {\em Phys. Rev. Lett.} {\bf 119} (2017), no.~12 121601,
  [\href{http://arxiv.org/abs/1609.00282}{{\tt arXiv:1609.00282}}].

\bibitem{Bagchi:2016bcd}
A.~Bagchi, R.~Basu, A.~Kakkar, and A.~Mehra, {\it {Flat Holography: Aspects of
  the dual field theory}},  {\em JHEP} {\bf 12} (2016) 147,
  [\href{http://arxiv.org/abs/1609.06203}{{\tt arXiv:1609.06203}}].

\bibitem{Cheung:2016iub}
C.~Cheung, A.~de~la Fuente, and R.~Sundrum, {\it {4D scattering amplitudes and
  asymptotic symmetries from 2D CFT}},  {\em JHEP} {\bf 01} (2017) 112,
  [\href{http://arxiv.org/abs/1609.00732}{{\tt arXiv:1609.00732}}].

\bibitem{Pasterski:2016qvg}
S.~Pasterski, S.-H. Shao, and A.~Strominger, {\it {Flat Space Amplitudes and
  Conformal Symmetry of the Celestial Sphere}},  {\em Phys. Rev. D} {\bf 96}
  (2017), no.~6 065026, [\href{http://arxiv.org/abs/1701.00049}{{\tt
  arXiv:1701.00049}}].

\bibitem{Pasterski:2017ylz}
S.~Pasterski, S.-H. Shao, and A.~Strominger, {\it {Gluon Amplitudes as 2d
  Conformal Correlators}},  {\em Phys. Rev. D} {\bf 96} (2017), no.~8 085006,
  [\href{http://arxiv.org/abs/1706.03917}{{\tt arXiv:1706.03917}}].

\bibitem{Cardona:2017keg}
C.~Cardona and Y.-t. Huang, {\it {S-matrix singularities and CFT correlation
  functions}},  {\em JHEP} {\bf 08} (2017) 133,
  [\href{http://arxiv.org/abs/1702.03283}{{\tt arXiv:1702.03283}}].

\bibitem{Ball:2019atb}
A.~Ball, E.~Himwich, S.~A. Narayanan, S.~Pasterski, and A.~Strominger, {\it
  {Uplifting AdS$_{3}$/CFT$_{2}$ to flat space holography}},  {\em JHEP} {\bf
  08} (2019) 168, [\href{http://arxiv.org/abs/1905.09809}{{\tt
  arXiv:1905.09809}}].

\bibitem{Donnay:2018neh}
L.~Donnay, A.~Puhm, and A.~Strominger, {\it {Conformally Soft Photons and
  Gravitons}},  {\em JHEP} {\bf 01} (2019) 184,
  [\href{http://arxiv.org/abs/1810.05219}{{\tt arXiv:1810.05219}}].

\bibitem{Himwich:2019dug}
E.~Himwich and A.~Strominger, {\it {Celestial current algebra from
  Low\textquoteright{}s subleading soft theorem}},  {\em Phys. Rev. D} {\bf
  100} (2019), no.~6 065001, [\href{http://arxiv.org/abs/1901.01622}{{\tt
  arXiv:1901.01622}}].

\bibitem{Fan:2019emx}
W.~Fan, A.~Fotopoulos, and T.~R. Taylor, {\it {Soft Limits of Yang-Mills
  Amplitudes and Conformal Correlators}},  {\em JHEP} {\bf 05} (2019) 121,
  [\href{http://arxiv.org/abs/1903.01676}{{\tt arXiv:1903.01676}}].

\bibitem{Pate:2019mfs}
M.~Pate, A.-M. Raclariu, and A.~Strominger, {\it {Conformally Soft Theorem in
  Gauge Theory}},  {\em Phys. Rev. D} {\bf 100} (2019), no.~8 085017,
  [\href{http://arxiv.org/abs/1904.10831}{{\tt arXiv:1904.10831}}].

\bibitem{Adamo:2019ipt}
T.~Adamo, L.~Mason, and A.~Sharma, {\it {Celestial amplitudes and conformal
  soft theorems}},  {\em Class. Quant. Grav.} {\bf 36} (2019), no.~20 205018,
  [\href{http://arxiv.org/abs/1905.09224}{{\tt arXiv:1905.09224}}].

\bibitem{Nandan:2019jas}
D.~Nandan, A.~Schreiber, A.~Volovich, and M.~Zlotnikov, {\it {Celestial
  Amplitudes: Conformal Partial Waves and Soft Limits}},  {\em JHEP} {\bf 10}
  (2019) 018, [\href{http://arxiv.org/abs/1904.10940}{{\tt arXiv:1904.10940}}].

\bibitem{Puhm:2019zbl}
A.~Puhm, {\it {Conformally Soft Theorem in Gravity}},  {\em JHEP} {\bf 09}
  (2020) 130, [\href{http://arxiv.org/abs/1905.09799}{{\tt arXiv:1905.09799}}].

\bibitem{Guevara:2019ypd}
A.~Guevara, {\it {Notes on Conformal Soft Theorems and Recursion Relations in
  Gravity}},  \href{http://arxiv.org/abs/1906.07810}{{\tt arXiv:1906.07810}}.

\bibitem{Fotopoulos:2019vac}
A.~Fotopoulos, S.~Stieberger, T.~R. Taylor, and B.~Zhu, {\it {Extended BMS
  Algebra of Celestial CFT}},  {\em JHEP} {\bf 03} (2020) 130,
  [\href{http://arxiv.org/abs/1912.10973}{{\tt arXiv:1912.10973}}].

\bibitem{Fotopoulos:2020bqj}
A.~Fotopoulos, S.~Stieberger, T.~R. Taylor, and B.~Zhu, {\it {Extended Super
  BMS Algebra of Celestial CFT}},  {\em JHEP} {\bf 09} (2020) 198,
  [\href{http://arxiv.org/abs/2007.03785}{{\tt arXiv:2007.03785}}].

\bibitem{Himwich:2020rro}
E.~Himwich, S.~A. Narayanan, M.~Pate, N.~Paul, and A.~Strominger, {\it {The
  Soft $\mathcal{S}$-Matrix in Gravity}},  {\em JHEP} {\bf 09} (2020) 129,
  [\href{http://arxiv.org/abs/2005.13433}{{\tt arXiv:2005.13433}}].

\bibitem{Magnea:2021fvy}
L.~Magnea, {\it {Non-abelian infrared divergences on the celestial sphere}},
  {\em JHEP} {\bf 05} (2021) 282, [\href{http://arxiv.org/abs/2104.10254}{{\tt
  arXiv:2104.10254}}].

\bibitem{Gonzalez:2021dxw}
H.~A. Gonz\'alez and F.~Rojas, {\it {The structure of IR divergences in
  celestial gluon amplitudes}},  \href{http://arxiv.org/abs/2104.12979}{{\tt
  arXiv:2104.12979}}.

\bibitem{Arkani-Hamed:2020gyp}
N.~Arkani-Hamed, M.~Pate, A.-M. Raclariu, and A.~Strominger, {\it {Celestial
  Amplitudes from UV to IR}},  \href{http://arxiv.org/abs/2012.04208}{{\tt
  arXiv:2012.04208}}.

\bibitem{Casali:2020uvr}
E.~Casali and A.~Sharma, {\it {Celestial double copy from the worldsheet}},
  {\em JHEP} {\bf 05} (2021) 157, [\href{http://arxiv.org/abs/2011.10052}{{\tt
  arXiv:2011.10052}}].

\bibitem{Casali:2020vuy}
E.~Casali and A.~Puhm, {\it {Double Copy for Celestial Amplitudes}},  {\em
  Phys. Rev. Lett.} {\bf 126} (2021), no.~10 101602,
  [\href{http://arxiv.org/abs/2007.15027}{{\tt arXiv:2007.15027}}].

\bibitem{Schreiber_2018}
A.~Schreiber, A.~Volovich, and M.~Zlotnikov, {\it {Tree-level gluon amplitudes
  on the celestial sphere}},  {\em Phys. Lett. B} {\bf 781} (2018) 349--357,
  [\href{http://arxiv.org/abs/1711.08435}{{\tt arXiv:1711.08435}}].

\bibitem{Banerjee:2017jeg}
N.~Banerjee, S.~Banerjee, S.~Atul~Bhatkar, and S.~Jain, {\it {Conformal
  Structure of Massless Scalar Amplitudes Beyond Tree level}},  {\em JHEP} {\bf
  04} (2018) 039, [\href{http://arxiv.org/abs/1711.06690}{{\tt
  arXiv:1711.06690}}].

\bibitem{Albayrak:2020saa}
S.~Albayrak, C.~Chowdhury, and S.~Kharel, {\it {On loop celestial amplitudes
  for gauge theory and gravity}},  {\em Phys. Rev. D} {\bf 102} (2020) 126020,
  [\href{http://arxiv.org/abs/2007.09338}{{\tt arXiv:2007.09338}}].

\bibitem{Gonzalez:2020tpi}
H.~A. Gonz\'alez, A.~Puhm, and F.~Rojas, {\it {Loop corrections to celestial
  amplitudes}},  {\em Phys. Rev. D} {\bf 102} (2020), no.~12 126027,
  [\href{http://arxiv.org/abs/2009.07290}{{\tt arXiv:2009.07290}}].

\bibitem{Stieberger:2018edy}
S.~Stieberger and T.~R. Taylor, {\it {Strings on Celestial Sphere}},  {\em
  Nucl. Phys. B} {\bf 935} (2018) 388--411,
  [\href{http://arxiv.org/abs/1806.05688}{{\tt arXiv:1806.05688}}].

\bibitem{Banerjee:2020kaa}
S.~Banerjee, S.~Ghosh, and R.~Gonzo, {\it {BMS symmetry of celestial OPE}},
  {\em JHEP} {\bf 04} (2020) 130, [\href{http://arxiv.org/abs/2002.00975}{{\tt
  arXiv:2002.00975}}].

\bibitem{Banerjee:2020vnt}
S.~Banerjee and S.~Ghosh, {\it {MHV Gluon Scattering Amplitudes from Celestial
  Current Algebras}},  \href{http://arxiv.org/abs/2011.00017}{{\tt
  arXiv:2011.00017}}.

\bibitem{Banerjee:2020zlg}
S.~Banerjee, S.~Ghosh, and P.~Paul, {\it {MHV graviton scattering amplitudes
  and current algebra on the celestial sphere}},  {\em JHEP} {\bf 02} (2021)
  176, [\href{http://arxiv.org/abs/2008.04330}{{\tt arXiv:2008.04330}}].

\bibitem{Pate:2019lpp}
M.~Pate, A.-M. Raclariu, A.~Strominger, and E.~Y. Yuan, {\it {Celestial
  Operator Products of Gluons and Gravitons}},
  \href{http://arxiv.org/abs/1910.07424}{{\tt arXiv:1910.07424}}.

\bibitem{Fan:2021isc}
W.~Fan, A.~Fotopoulos, S.~Stieberger, T.~R. Taylor, and B.~Zhu, {\it {Conformal
  blocks from celestial gluon amplitudes}},  {\em JHEP} {\bf 05} (2021) 170,
  [\href{http://arxiv.org/abs/2103.04420}{{\tt arXiv:2103.04420}}].

\bibitem{Strominger:2021lvk}
A.~Strominger, {\it {w(1+infinity) and the Celestial Sphere}},
  \href{http://arxiv.org/abs/2105.14346}{{\tt arXiv:2105.14346}}.

\bibitem{Atanasov:2021cje}
A.~Atanasov, W.~Melton, A.-M. Raclariu, and A.~Strominger, {\it {Conformal
  Block Expansion in Celestial CFT}},
  \href{http://arxiv.org/abs/2104.13432}{{\tt arXiv:2104.13432}}.

\bibitem{Guevara:2021abz}
A.~Guevara, E.~Himwich, M.~Pate, and A.~Strominger, {\it {Holographic Symmetry
  Algebras for Gauge Theory and Gravity}},
  \href{http://arxiv.org/abs/2103.03961}{{\tt arXiv:2103.03961}}.

\bibitem{Crawley:2021ivb}
E.~Crawley, N.~Miller, S.~A. Narayanan, and A.~Strominger, {\it {State-Operator
  Correspondence in Celestial Conformal Field Theory}},
  \href{http://arxiv.org/abs/2105.00331}{{\tt arXiv:2105.00331}}.

\bibitem{Atanasov:2021oyu}
A.~Atanasov, A.~Ball, W.~Melton, A.-M. Raclariu, and A.~Strominger, {\it
  {$(2,2)$ Scattering and the Celestial Torus}},
  \href{http://arxiv.org/abs/2101.09591}{{\tt arXiv:2101.09591}}.

\bibitem{Brandhuber:2021nez}
A.~Brandhuber, G.~R. Brown, J.~Gowdy, B.~Spence, and G.~Travaglini, {\it
  {Celestial Superamplitudes}},  \href{http://arxiv.org/abs/2105.10263}{{\tt
  arXiv:2105.10263}}.

\bibitem{Jiang:2021xzy}
H.~Jiang, {\it {Celestial superamplitude in $\mathcal N=4$ SYM theory}},
  \href{http://arxiv.org/abs/2105.10269}{{\tt arXiv:2105.10269}}.

\bibitem{Drummond:2008vq}
J.~M. Drummond, J.~Henn, G.~P. Korchemsky, and E.~Sokatchev, {\it {Dual
  superconformal symmetry of scattering amplitudes in N=4 super-Yang-Mills
  theory}},  {\em Nucl. Phys. B} {\bf 828} (2010) 317--374,
  [\href{http://arxiv.org/abs/0807.1095}{{\tt arXiv:0807.1095}}].

\bibitem{aomoto}
K.~Aomoto and M.~Kita, {\em Theory of Hypergeometric Functions}.
\newblock Springer Japan, Reading, Massachusetts, 2011.

\bibitem{celestiahedron}
Y.~Hu, L.~Ren, M.~Spradlin, A.~Yelleshpur~Srikant, and A.~Volovich, {\it {The
  Celestiahedron}},  {\em work in progress} (2021)
  [\href{http://arxiv.org/abs/yymm.nnnnn}{{\tt arXiv:yymm.nnnnn}}].

\bibitem{Elvang:2015rqa}
H.~Elvang and Y.-t. Huang, {\em {Scattering Amplitudes in Gauge Theory and
  Gravity}}.
\newblock Cambridge University Press, 4, 2015.

\bibitem{Henn:2014yza}
J.~M. Henn and J.~C. Plefka, {\em {Scattering Amplitudes in Gauge Theories}},
  vol.~883.
\newblock Springer, Berlin, 2014.

\bibitem{Drummond:2009fd}
J.~M. Drummond, J.~M. Henn, and J.~Plefka, {\it {Yangian symmetry of scattering
  amplitudes in N=4 super Yang-Mills theory}},  {\em JHEP} {\bf 05} (2009) 046,
  [\href{http://arxiv.org/abs/0902.2987}{{\tt arXiv:0902.2987}}].

\bibitem{Law:2019glh}
Y.~T.~A. Law and M.~Zlotnikov, {\it {Poincar\'e constraints on celestial
  amplitudes}},  {\em JHEP} {\bf 03} (2020) 085,
  [\href{http://arxiv.org/abs/1910.04356}{{\tt arXiv:1910.04356}}]. [Erratum:
  JHEP 04, 202 (2020)].

\bibitem{Stieberger:2018onx}
S.~Stieberger and T.~R. Taylor, {\it {Symmetries of Celestial Amplitudes}},
  {\em Phys. Lett. B} {\bf 793} (2019) 141--143,
  [\href{http://arxiv.org/abs/1812.01080}{{\tt arXiv:1812.01080}}].

\bibitem{Abe}
Y.~Abe, {\it A note on generalized hypergeometric functions, kz solutions, and
  gluon amplitudes},  {\em Nuclear Physics B} {\bf 907} (Jun, 2016) 107–153.

\end{thebibliography}\endgroup


\providecommand{\href}[2]{#2}\begingroup\raggedright\endgroup
\bibliographystyle{JHEP}

\end{document}